\RequirePackage{fix-cm}

\documentclass[smallextended]{svjour3}       
\smartqed  
\usepackage{graphicx}

\usepackage[english]{babel}
\usepackage[utf8]{inputenc}
\usepackage{amsmath}

\usepackage{geometry}
\usepackage[most]{tcolorbox}
\usepackage{fancyhdr}

\usepackage{graphicx}
\usepackage{xcolor}
\usepackage{tikz}
\usepackage{pgfgantt}
\usepackage{float}
\usepackage{enumitem}
\usepackage{booktabs}
\usepackage{tabularx}
\usepackage{tabularray}
\usepackage{ragged2e}

\usepackage{natbib}
\usepackage{eurosym}
\usepackage{url}

\usepackage[hidelinks]{hyperref}

\usepackage{listings}
\lstdefinelanguage{json}{
  basicstyle=\ttfamily\small,
  numbers=left,
  numberstyle=\tiny,
  stepnumber=1,
  numbersep=6pt,
  showstringspaces=false,
  breaklines=true,
  frame=single,
  backgroundcolor=\color{gray!5},
  stringstyle=\color{blue},
  keywordstyle=\color{black},
  morekeywords={true,false,null}
}

\begin{document}

\title{Detect–Repair–Verify for Securing LLM-Generated Code: A Multi-Language Empirical Study
}


\author{Cheng Cheng 
}

\institute{Cheng Cheng \at
              Department of Computer Science and \\ Software Engineering, Concordia University,\\Montreal, Canada\\
              \email{cheng.cheng.20171@mail.concordia.ca}           
}
\date{Received: date / Accepted: date}

\maketitle

\begin{abstract}
Large language models are increasingly used to produce runnable software. In practice, security is often addressed through a Detect--Repair--Verify (DRV) loop that detects issues, applies fixes, and verifies the result. This work studies such a workflow for project-level artifacts and addresses four gaps: L1, the lack of project-level benchmarks with executable function and security tests; L2, limited evidence on pipeline-level effectiveness beyond studying detection or repair alone; L3, unclear reliability of detection reports as repair guidance; and L4, uncertain repair trustworthiness and side effects under verification. A new benchmark dataset\footnote{\url{https://github.com/Hahappyppy2024/EmpricalVDR}} is introduced, consisting of runnable web-application projects paired with functional tests and targeted security tests, and supporting three prompt granularities at the project, requirement, and function level. The evaluation compares generation-only, single-pass DRV, and bounded iterative DRV variants under comparable budget constraints. Outcomes are measured by secure and correct yield using test-grounded verification, and intermediate artifacts are analyzed to assess report actionability and post-repair failure modes such as regressions, semantic drift, and newly introduced security issues.

\end{abstract}

\section{Introduction}
\label{intro}

Large language models are now part of everyday software development. They can translate natural-language requirements into runnable code and speed up prototyping and implementation across a wide range of tasks, from writing individual functions to assembling complete, project-level artifacts. In this setting, security should be understood through the lens of vulnerability management, which treats software protection as a continuous process rather than a single step at coding time. A typical vulnerability management workflow starts with identifying security-relevant issues during development and maintenance, then detecting potential vulnerabilities in code or running systems, analyzing and localizing the root cause to specific components or code regions, and prioritizing what to handle first based on risk and impact. It then moves to remediation, where developers repair vulnerabilities by applying patches or refactoring insecure logic, followed by verification to ensure the fix is effective and does not break intended behavior. Finally, the workflow includes fix identification and tracking, which links patches to vulnerabilities for auditing, release management, and downstream propagation, so that related systems and versions can be updated consistently. When LLMs participate in software creation, they enter this same lifecycle, and securing LLM-generated code requires the same end-to-end management across detection, repair, and verification.

Building on this lifecycle perspective, when code is produced by large language models, vulnerability management can be operationalized as a structured Detect–Repair–Verify (DRV) workflow for securing LLM-generated project artifacts. After an initial generation step, potential vulnerabilities are identified through automated detection tools or model-based analysis. The reported issues then guide targeted repair actions, which may also be assisted by LLMs. Finally, the repaired artifact is re-evaluated through security checks and functional tests to confirm that vulnerabilities are mitigated and intended behavior is preserved. This Detect–Repair–Verify workflow reflects how vulnerability management is enacted in practice when LLM-generated artifacts enter development pipelines, and it provides a concrete basis for multi-language empirical evaluation under comparable verification protocols.

Despite the rapid uptake of LLMs in programming workflows, existing results do not yet establish how well an end-to-end security-hardening pipeline works for LLM-generated projects. Much of the literature evaluates vulnerability detection or vulnerability repair in isolation, rather than the composed generate--detect--repair--verify workflow that developers would execute in practice. This gap matters because detection outputs can be unstable and sensitive to superficial code variations, undermining their reliability as repair guidance and leading to missed issues or spurious alarms \citep{ullah2024sp_secLLMHolmes}. In parallel, repair-oriented studies increasingly stress the need for post-repair validation—such as rerunning security checks and functional tests—to confirm mitigation and to rule out regressions and unintended side effects \citep{wang2025naacl_cvebench,kim2025usenix_san2patch,zhang2024nist_vulrepair,weI2025arxiv_patcheval}. Together, these observations motivate a systematic evaluation of the workflow as a whole, focusing on pipeline-level effectiveness, the usefulness of detection reports for guiding repair, and the verified quality of repaired outputs.

\textbf{Limitation 1 (Lack of project-level, test-grounded benchmarks).}
Most existing evaluations rely on simplified artifacts (e.g., isolated functions or snippets) or datasets without executable, security-relevant test oracles. This limits the ability to study LLM-generated \emph{projects} under realistic verification conditions, where functional tests and security tests jointly determine whether an artifact is both correct and secure. A project-level benchmark spanning multiple language ecosystems and paired with functional tests and security tests therefore remains underrepresented.

\textbf{Limitation 2 (Unverified pipeline-level effectiveness).}
Existing studies typically assess vulnerability detection or vulnerability repair in isolation. Consequently, evidence is still limited on whether an end-to-end generate--detect--repair--verify workflow consistently improves the secure-and-correct yield of \emph{LLM-generated projects} over generation-only baselines.

\textbf{Limitation 3 (Unclear detection reliability as repair guidance).}
Detection reports on LLM-generated code can be unstable and uneven in evidential quality. When used as repair guidance, missed findings may leave vulnerabilities unaddressed, while spurious findings can prompt unnecessary or misdirected edits.

\textbf{Limitation 4 (Uncertain repair trustworthiness and side effects).}
LLM-based repair may remove reported issues, but it can also introduce regressions, semantic drift, or new security flaws. As a result, the trustworthiness of ``fixed'' project-level outputs under verification, and the failure modes that dominate in practice, remain insufficiently characterized.

These limitations motivate a systematic empirical study of LLM-based detect-and-repair workflows as an integrated process, focusing on when they improve security and correctness and why they succeed or fail. Accordingly, this proposal is guided by the following research questions.

Accordingly, the study is structured around the following research questions:

\noindent\textbf{RQ1 (Pipeline-level effectiveness).}
Under comparable budget constraints, does a bounded iterative detect--repair--verify workflow with test-grounded feedback improve the secure-and-correct yield of project-level artifacts relative to generation-only and single-pass detect--repair baselines, and how does the effect vary with prompt granularity?

\noindent\textbf{RQ2 (Detection reliability as repair guidance).}
How reliable and repair-actionable are LLM-generated detection reports when applied to project-level artifacts?

\noindent\textbf{RQ3 (Repair trustworthiness and dominant failure modes).}
To what extent can LLM-based repair mitigate reported vulnerabilities in iterative workflows without introducing functional regressions, and which failure modes dominate under verification?

To answer these questions, the study evaluates generation-only, single-pass detect--repair, and bounded iterative detect--repair--verify variants under comparable budget constraints with test-grounded feedback. A project-level benchmark dataset is introduced to enable end-to-end measurement at the project granularity. The dataset comprises runnable web-application projects, each paired with executable functional tests and targeted security tests, and it supports three prompt granularities (project-, requirement-, and function-level) to control generation and iteration scope. The experimental design compares these workflow variants across granularities, treating each project-level artifact as the unit of analysis. Outcomes are quantified using secure-and-correct yield: correctness is assessed by functional tests, while security is assessed by security tests, complemented by post-repair re-detection. Beyond final outcomes, intermediate artifacts are analyzed to assess the reliability and repair-actionability of detection reports (RQ2) and to characterize repair side effects under verification (RQ3), including functional regressions, semantic drift, and newly introduced security issues. This combined outcome-level evaluation and process-level tracing supports both effectiveness assessment (RQ1) and failure-mode attribution.

The remainder of this paper is organized as follows. Section~\ref{related} reviews prior work on LLM-assisted vulnerability detection, automated repair, and verification-oriented evaluation. Section~\ref{methodology} describes the experimental design, including the studied workflows and baselines, budget and iteration constraints, prompt granularities, subject systems, and the functional and security test suites used for verification. Section~\ref{Results} reports the empirical results for RQ1--RQ3, covering pipeline-level secure-and-correct yield, detection reliability as repair guidance, and repair trustworthiness with dominant failure modes. Section~\ref{disscusion} discusses the main findings and their implications for LLM-based vulnerability management workflows. Section~\ref{threats} discusses threats to validity and mitigating measures. Section~\ref{conclusion} concludes with key findings and implications for LLM-based vulnerability management workflows.

\section{Related Work}
\label{related}

Large language models (LLMs) are increasingly used in security-relevant software engineering tasks. Existing work spans four closely related directions: secure code generation, vulnerability detection and localization, automated vulnerability repair, and prompting or workflow strategies that steer model behavior. This study connects these directions by examining detect--repair--verify workflows with test-grounded evidence across multiple language implementations and prompt granularities. The rest of this section reviews prior work in each area and clarifies how it relates to the proposed evaluation setting.

\subsection{Secure Code Generation by LLMs}
A growing body of work investigates whether LLM-generated code is secure by default and how to improve its security properties. The literature increasingly moves beyond functional correctness to evaluation protocols that jointly assess correctness and security using testable evidence. SecurityEval provides CWE-aligned prompts and examples to probe vulnerability-prone generations, while SALLM further systematizes the evaluation pipeline and metrics to reduce reliance on ad hoc rules or subjective judgments~\citep{siddiq2022securityeval,siddiq2024sallm}. Along this evaluation-centric line, Cheng et al. propose benchmarking security-relevant behaviors in LLM-based code generation~\citep{cheng2024benchmarking}, and CFCEval offers a more structured set of security-oriented criteria with evidence-based checks for assessing whether generated implementations satisfy security expectations~\citep{cheng2025cfceval}. Beyond benchmarks, recent efforts incorporate security into model training and generation-time control: SafeCoder improves security via instruction tuning~\citep{he2024safecoder}, whereas SCodeGen introduces constrained decoding to steer outputs toward trustworthy implementations in real time~\citep{SCodeGen2025Qu}. Finally, newer benchmarks such as BaxBench and CWEval push toward more realistic and reproducible assessment by combining executable tests with exploit-oriented or outcome-driven verification, highlighting the prevalence of code that is correct yet insecure~\citep{vero2025baxbench,peng2025cweval}. Complementing these benchmark-driven studies, empirical evidence from S\&P/CCS shows that AI coding assistants measurably affect developers' security behavior and the security quality of their outputs, suggesting that security cannot be assumed by default~\citep{pearce2022copilot,perry2023insecureusers}. Collectively, these findings motivate security-aware generation and rigorous, test-grounded evaluation, aligning with the need for evidence-based workflows when studying detect--repair--verify pipelines.

\subsection{Vulnerability Detection}

Prior work on vulnerability detection can be broadly categorized into traditional static analysis, dynamic analysis, learning-based detection, and recent LLM-based approaches, which differ in their reliance on program semantics, runtime behavior, and data-driven modeling.

\paragraph{Traditional static analysis.}
Static vulnerability detection analyzes program code without execution and relies on explicit program semantics such as data flow and taint propagation. STASE~\citep{stase2024ase} exemplifies this line by applying static taint analysis to identify configuration-related vulnerabilities, offering interpretable and scalable detection while remaining sensitive to analysis precision and contextual complexity.

\paragraph{Dynamic analysis.}
Dynamic approaches identify vulnerabilities by observing runtime behavior. DeFiWarder~\citep{defiwarder2023ase} illustrates this paradigm by detecting token-level vulnerabilities in DeFi applications through runtime semantics, though dynamic techniques are inherently limited by execution coverage and input exploration.

\paragraph{Learning-based detection.}
\paragraph{Learning-based detection.}
Learning-based methods infer vulnerability patterns from data using learned representations of programs, commonly built on structured graphs. MVD leverages flow-sensitive graph neural networks for memory vulnerability detection \citep{mvd2022icse}. Later work adds explanation mechanisms to improve interpretability and support analyst-facing use, as exemplified by Coca and CFExplainer \citep{coca2024icse,cfexplainer2024acm}. In the context of detect--repair workflows, an additional consideration is the quality of detection outputs as actionable guidance, including the stability of findings, the specificity of evidence, and the precision of localization.

\paragraph{Vulnerability Localization}
Vulnerability localization identifies the precise code locations responsible for security vulnerabilities—typically at the function-, statement-, or line-level—and serves as a critical bridge between vulnerability detection and subsequent remediation. Compared with coarse-grained detection, localization places stronger requirements on semantic understanding and granularity, since it must not only determine whether a vulnerability exists but also pinpoint where it manifests in the program. Early learning-based approaches model vulnerable code patterns directly for fine-grained localization: VulDeeLocator performs statement-level localization with neural models over code representations, showing that learned features can move beyond file- or function-level detection~\citep{li2022vuldeelocator}. Building on this direction, VulTeller couples localization with vulnerability description generation, enabling models to both highlight vulnerable regions and produce natural-language explanations of the underlying issues~\citep{zhang2023vulteller}. More recent studies explore large language models to improve localization by leveraging stronger contextual reasoning and configuration choices: ENVUL examines LLM-based localization and studies how task-specific tuning interacts with prompting strategies to yield effective setups~\citep{tian2025envul}. In parallel, VulPCL jointly tackles vulnerability prediction, categorization, and localization, integrating localization with complementary analysis tasks to provide more actionable code-level outputs~\citep{liu2024vulpcl}. Overall, the literature trends toward more context-aware and holistic localization methods that connect detection signals to actionable remediation guidance.

\subsection{Vulnerability Repair}

Automated vulnerability repair aims to generate or recommend code changes that eliminate security vulnerabilities while preserving program functionality. Early research in this area spans a wide range of techniques, including rule-based patch generation, pattern mining, and learning-based approaches. A comprehensive overview of this landscape is provided by the SoK~\citep{Hu0SGZXY025} on automated vulnerability repair, which systematically reviews existing methods, tools, datasets, and evaluation practices, and highlights common challenges such as patch correctness, generalization, and evaluation reliability.

\subsubsection{Pattern- and learning-based vulnerability repair.}
Learning-based repair approaches have gained attention due to their ability to generalize repair patterns from data. VulRepair~\citep{fu2022vulrepair} represents a representative sequence-to-sequence approach that formulates vulnerability repair as a code translation problem using a T5-based model, demonstrating the feasibility of neural models for generating vulnerability-fixing patches. Complementary to direct patch generation, VulMatch~\citep{cao2025vulmatch} focuses on extracting and matching repair patterns from historical fixes, leveraging recurring fix templates to enhance repair effectiveness and interpretability.

\subsubsection{LLM-based Vulnerability Repair}
\paragraph{Early exploration of LLM capabilities for repair.}
In parallel, early studies have explored the repair capabilities of large language models without task-specific fine-tuning. Pearce et al.~\citeyearpar{pearce2021zeroshot} examine zero-shot vulnerability repair with large language models, establishing an important baseline and revealing both the potential and limitations of LLMs when applied to vulnerability repair without explicit training signals. Together, these works illustrate the evolution of automated vulnerability repair from pattern- and model-driven approaches toward more flexible, data-driven repair paradigms.

\paragraph{Empirical understanding of LLM-based repair.}
More recent work explicitly investigates large language models as a central component in automated vulnerability repair. VRpilot~\citep{kulsum2024vrpilot} presents an in-depth case study of LLM-based vulnerability repair, analyzing how reasoning strategies and patch validation feedback influence repair performance and reliability. Rather than proposing a single end-to-end solution, this work provides empirical insights into the behavior of LLMs in vulnerability repair settings.

\paragraph{Prompt-driven and iterative LLM repair.}
Subsequent studies aim to enhance LLM-based repair through improved prompting and adaptation strategies. De-Fitero-Dominguez et al.~\citeyearpar{defiterodominguez2024enhancedavr} propose an approach that leverages large language models for automated code vulnerability repair, demonstrating that careful prompt design and task formulation can substantially affect repair outcomes. LLM4CVE~\citep{fakih2025llm4cve} further extends this direction by enabling iterative vulnerability repair, where LLMs interact with vulnerability information and intermediate feedback to progressively refine repair candidates.

\paragraph{CWE-guided and fine-grained control of LLM repair.}
Beyond prompting and iteration, recent work explores more fine-grained control of LLM behavior for vulnerability repair. The fine-grained cue optimization approach combines localized cues with sensitive fine-tuning strategies to guide LLMs toward more accurate and targeted repairs, aiming to improve patch correctness and stability across different vulnerability types
~\citep{zhang2025cueopt}. 
Overall, these studies indicate a shift from one-shot patch generation toward iterative, feedback-driven, and cue-aware LLM-based vulnerability repair frameworks.

\subsection{Prompting Engineering}

Early studies on prompting established its effectiveness as an alternative to fine-tuning for controlling model behavior. A comprehensive overview of prompt-based learning is provided by Liu et al.~\citeyearpar{liu2023pretrainpromptpredict}, who systematically survey prompting methods and categorize them according to their design principles and application scenarios.

A major line of work focuses on eliciting multi-step reasoning through structured prompts. Chain-of-thought~ \citep{wei2022cot} prompting demonstrates that explicitly prompting intermediate reasoning steps can significantly improve performance on reasoning-intensive tasks. Subsequent work shows that large language models can perform such reasoning even in zero-shot settings by using simple trigger phrases \citep{kojima2022zeroshotcot}. To improve the robustness of reasoning, self-consistency aggregates multiple reasoning paths and selects consistent outcomes \citep{wang2023selfconsistency}, while least-to-most prompting decomposes complex problems into simpler subproblems to enable progressive reasoning \citep{zhou2023leasttomost}. Tree-of-thoughts~\citep{yao2023tot} further generalizes these ideas by framing reasoning as a structured search over multiple intermediate thought trajectories.

Beyond reasoning elicitation, recent studies explore prompt optimization and reuse. Wan et al.~\citeyearpar{wan2024teachbetter} investigate the relative roles of instructions and exemplars in automatic prompt optimization, while localized zeroth-order prompt optimization proposes black-box optimization strategies for improving prompts without gradient access \citep{hu2024zopo}. Buffer-of-thoughts~\citep{yang2024bot} introduces reusable reasoning templates to enhance thought-augmented reasoning across tasks , and dynamic rewarding with prompt optimization enables tuning-free self-alignment by adjusting prompts based on reward feedback \citep{singla2024drpo}. Complementary to these approaches, DSPy~\citep{khattab2024dspy} treats prompting as a programmable and optimizable pipeline, compiling declarative specifications into self-improving prompt workflows .

\section{Methodology}
\label{methodology}

\begin{figure}
    \centering
    \includegraphics[width=1\linewidth]{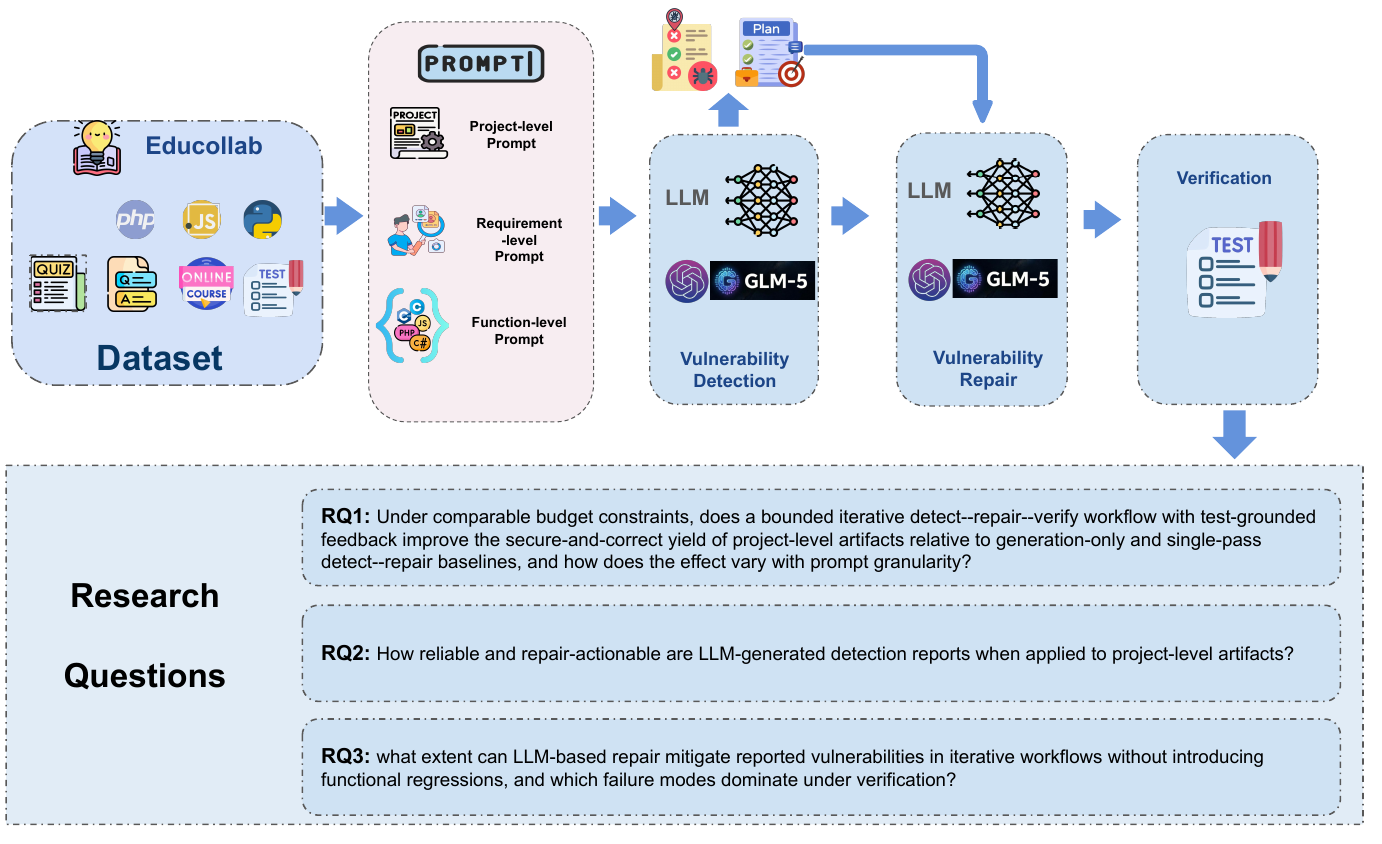}
    \caption{Overview of the experimental pipeline. Starting from the EduCollab benchmark (PHP/JS/Python), artifacts are constructed under three prompt granularities (project-, requirement-, and function-level) and processed through a detect--repair--verify workflow. Verification is performed via the test suite, whose outcomes provide grounded feedback to drive bounded iterations and enable measurement of RQ1--RQ3.}
    \label{fig:methodology}
\end{figure}

This section describes the study methodology as an end-to-end pipeline (Fig.~\ref{fig:methodology}). A multi-language EduCollab benchmark is constructed with paired functional and security test suites. Two LLMs are then applied to vulnerability detection and repair under different prompt granularities and workflow conditions. Each proposed patch is finally validated by rerunning the corresponding security target together with the full functional suite. The verified outcomes are recorded per target and summarized per model--language--workflow setting to support the analyses in RQ1--RQ3.

\subsection{Dataset Construction}
The dataset consists of three independently implemented web applications, EduCollab-PHP, EduCollab-JS, and EduCollab-Python, each following an MVC architecture. All three applications are course-centric and multi-tenant, using \textit{Course} as the tenant boundary. Although the implementations differ in language and framework, they follow the same functional specification, which enables controlled cross-language comparisons under matched application semantics. The dataset and accompanying artifacts are available in a public repository.\footnote{\url{https://github.com/Hahappyppy2024/EmpricalVDR}}

Dataset construction follows five stages. First, a shared functional specification is defined to standardize application scope and expected user workflows across implementations. Second, three language-specific projects are implemented to match this blueprint, with consistent role boundaries and tenant isolation semantics. Third, the supported use cases are mapped to the OWASP Top~10:2025 (\texttt{R01--R10}) risk taxonomy to characterize the intended security exposure of each functional surface. Fourth, two complementary test suites are developed for each project. The functional tests validate intended workflows under benign inputs, while the security tests probe Top~10-relevant attack surfaces and provide observable failure signals for verification. Fifth, three prompt granularities are prepared for LLM interaction at the project level, requirement level, and function level. This supports evaluating detect, repair, and verify workflows under comparable budgets while varying how much context and locality is provided to the model. Overall, the staged design keeps application semantics aligned across languages and supports reproducible, test-grounded evaluation of secure-and-correct yield across prompt granularities.

\subsubsection{Stage~1: Functional Specification}

The dataset is built around a single shared functional specification to support controlled comparison across languages. Three EduCollab implementations in PHP, JavaScript, and Python are developed against the same blueprint, so that differences observed in detection, repair, and verification reflect workflow behavior and language or runtime effects rather than differences in application scope. The specification captures representative workflows of a course-centric collaboration platform. It includes authentication and session management, course and membership management with role-based access control, token-based invite joining, content posting and search, file submission and resource distribution, and administrative functions such as auditing and data import and export.

Table~\ref{tab:functional_features_usecases} lists the concrete use cases instantiated in the projects. Several use cases are security-relevant by design, including rendering and search surfaces, ZIP and CSV import and download, path handling, misconfiguration checks, audit logging, and exceptional-condition handling. These features provide realistic attack surfaces commonly targeted in web application security evaluation. All three implementations expose the same set of use cases and are validated with a shared functional test suite to establish semantic equivalence before security testing and the subsequent LLM-driven detect, repair, and verify experiments.

\begin{table}
\centering
\caption{Functional use cases instantiated in EduCollab (shared across PHP/JS/Python).}
\label{tab:functional_features_usecases}
\begin{tblr}{
  row{1} = {font=\bfseries},
  hlines = {dotted},
  hline{1,16} = {-}{solid,0.08em},
  hline{2} = {-}{solid,0.05em},
}
ID  & Use case                                                                           \\
U1  & Course and membership management (RBAC).                                           \\
U2  & Role update (member role change).                                                  \\
U3  & Security misconfiguration checks (cookie flags, \texttt{nosniff}, debug leakage).  \\
U4  & Invite-link join (token-based enrollment).                                         \\
U5  & Post/comment and search (keyword, sort).                                           \\
U6  & Stored/reflected rendering surfaces (XSS-relevant outputs).                        \\
U7  & ZIP/CSV import and download (CSV formula surface).                                 \\
U8  & Self-escalation via role change (assistant promotes self to admin).                \\
U9  & Authentication and session lifecycle (login/logout).                               \\
U10 & Assignment submission with upload (file handling).                                 \\
U11 & Upload/download with path-handling constraints (integrity-relevant surfaces).      \\
U12 & ZIP/CSV import and download (integrity checks).                                    \\
U13 & Admin audit-log view (\texttt{/admin/audit}); missing logging for critical events. \\
U14 & Exceptional-condition handling (malformed inputs; error leakage surfaces).         
\end{tblr}
\end{table}

\subsubsection{Stage~2: Cross-language Implementation}
Based on the shared functional specification in Stage~1, three EduCollab variants are implemented in PHP, JavaScript using Node.js and Express, and Python using Flask. Each variant follows an MVC-style organization that separates routing and controllers, data access, and server-rendered views, and uses SQLite for persistence. Keeping the scope and data model aligned across languages makes it possible to attribute observed differences in detection, repair, and verification to language and framework factors rather than to missing or inconsistent functionality.

Comparability is further strengthened through shared tenancy and authorization semantics. A course serves as the tenant boundary for data isolation, and the same role-based access control rules apply to the same resource types, including courses, memberships, posts, uploads, assignments, and submissions, across Admin, Teacher, Assistant, Student, and Unauthenticated roles. When applicable, each use case is exposed through both Web UI workflows and corresponding \texttt{/api/*} endpoints. This allows the same functional checks and security checks to run in both interactive and programmatic settings.

\subsubsection{Stage~3: OWASP Top~10:2025 Risk Mapping}

This stage describes the dataset's security exposure by mapping implemented use cases to the OWASP Top~10:2025~\citeyearpar{owasp_top10_2025} risk taxonomy (\texttt{R01--R10}). The intent is not to provide exhaustive coverage of web security issues, but to align EduCollab with widely used risk families so that evaluation and reporting remain comparable across languages. A course-centric collaboration platform provides an appropriate setting because it includes common security-relevant surfaces such as authentication and session management, role-based authorization, administrative operations, user-generated content rendering, and data and file import and export.

Table~\ref{tab:top10_summary_ranked} summarizes the Top~10 risk categories used in this study. Each entry lists the category name, its identifier (\texttt{R01--R10}), and an operational interpretation that informs test design and result reporting. The original OWASP ranking is kept for readability, while the taxonomy identifiers serve as the primary references in the remainder of the paper.

Table~\ref{tab:feature_rank_2025_sorted} links the dataset's functional use cases to the Top~10 categories. Use cases are grouped by risk identifier and ordered by ascending rank to match Table~\ref{tab:top10_summary_ranked}. Multiple use cases can map to the same category because a single risk family may appear in different parts of the application, such as authorization checks in both membership management and role updates. The mapping covers nine categories (\texttt{R01}, \texttt{R02}, \texttt{R04--R10}). \texttt{R03}, which concerns software supply chain failures, is primarily driven by dependency choices and build or deployment processes, and is therefore outside the scope of in-application use cases included in this dataset.

\begin{table}
\centering
\caption{2025 Top-10 web application security risks.}
\label{tab:top10_summary_ranked}
\begin{tblr}{
  row{1} = {c,font=\bfseries},
  cell{2}{1} = {c},
  cell{3}{1} = {c},
  cell{4}{1} = {c},
  cell{5}{1} = {c},
  cell{6}{1} = {c},
  cell{7}{1} = {c},
  cell{8}{1} = {c},
  cell{9}{1} = {c},
  cell{10}{1} = {c},
  cell{11}{1} = {c},
  hlines = {dotted},
  hline{1-2,12} = {-}{solid},
}
Rank & Category (ID)                                             & What it means                                                                                                 \\
1    & Broken Access Control (R01)                               & {Authorization checks are missing or bypassable,\\so users can access data/actions beyond their permissions.} \\
2    & Security Misconfiguration (R02)                           & {The app, server, framework, or cloud settings~\\are insecure or left in unsafe defaults.}                    \\
3    & Software Supply Chain Failures (R03)                      & {Risks come from dependencies, build artifacts,\\~or delivery pipelines rather than your own code.}           \\
4    & Cryptographic Failures (R04)                              & {Data is not properly protected because cryptography\\~is missing, weak, or misused.}                         \\
5    & Injection (R05)                                           & {Untrusted input is interpreted as code/commands/queries\\~by an interpreter or engine.}                      \\
6    & Insecure Design (R06)                                     & {The system’s design lacks necessary security controls~\\or abuse-resistance from the start.}                 \\
7    & Authentication Failures (R07)                             & {Identity and session mechanisms are weak, allowing~\\account takeover or session abuse.}                     \\
8    & Software or Data Integrity Failures (R08)                 & {The app trusts code/data that can be tampered with,\\leading to unsafe updates or corrupted flows.}          \\
9    & Security Logging \textbackslash{} Alerting Failures (R09) & {Security-relevant events are not monitored well,~\\delaying detection and response.}                         \\
10   & Mishandling of Exceptional Conditions (R10)               & {Error/exception paths are unsafe and can leak\\~information, bypass checks, or fail open.}                   
\end{tblr}
\end{table}

\begin{table}[t]
\centering
\caption{Functional features (use cases) aligned with the ranked Top-10 risk list (R01--R10), sorted by ascending rank. The table covers 9 of the 10 risk categories (R01, R02, R04--R10); R03 (software supply chain failures) is treated as a process-level risk and thus not represented as an in-app use case.}
\label{tab:feature_rank_2025_sorted}
\begin{tblr}{
  row{1} = {c,font=\bfseries},
  cell{2}{1} = {c},
  cell{3}{1} = {c},
  cell{4}{1} = {c},
  cell{5}{1} = {c},
  cell{6}{1} = {c},
  cell{7}{1} = {c},
  cell{8}{1} = {c},
  cell{9}{1} = {c},
  cell{10}{1} = {c},
  cell{11}{1} = {c},
  cell{12}{1} = {c},
  cell{13}{1} = {c},
  cell{14}{1} = {c},
  cell{15}{1} = {c},
  hlines = {dotted},
  hline{1-2,16} = {-}{solid},
}
\textbf{Rank ID} & \textbf{Functional feature (use case)} \\
R01 & Course \textbackslash{} Membership (RBAC) \\
R01 & Role Update (member role change) \\

R02 & Security Misconfiguration (cookie flags / nosniff / debug leakage) \\

R04 & Invite Link Join (token-based join) \\

R05 & Post/Comment \textbackslash{} Search (keyword/sort) \\
R05 & Stored/Reflected Rendering (XSS surface) \\
R05 & ZIP/CSV Import \textbackslash{} Download (CSV formula) \\

R06 & Self-escalation via Role Change (assistant can promote self to admin) \\

R07 & Auth \textbackslash{} Session (login/logout) \\

R08 & Assignment Submission Upload (file handling) \\
R08 & Upload/Download \textbackslash{} Path Handling (integrity surfaces) \\
R08 & ZIP/CSV Import \textbackslash{} Download (integrity) \\

R09 & Admin Audit Log View (/admin/audit) with missing logging for critical events \\

R10 & Exceptional Conditions (malformed inputs / error leakage) \\
\end{tblr}
\end{table}

\subsubsection{Stage~4: Functional and Security Test Suites}

This stage turns the dataset into an executable benchmark by pairing each implementation with two complementary test suites. The function test suite checks functional correctness, while the security test suite checks security expectations at high risk surfaces. Together, they provide test grounded evidence for both correctness and security during evaluation.

\paragraph{Functional correctness tests.}
functional tests exercise the expected end to end workflows of a course centric collaboration system, including authentication and session handling, course creation and membership management, invite based joining, posting and search, assignment submission with uploads, and data import and export. Tests are executed under appropriate authorization contexts so that intended actions succeed for authorized roles and are rejected for unauthorized roles, as specified in the shared functional blueprint.

\paragraph{Security tests.}
Security tests probe Top~10 related attack surfaces and are designed to yield clear pass or fail signals. The suite targets representative vulnerability families, including broken access control such as cross tenant access and IDOR style behaviors, injection style input handling issues, unsafe file and archive processing, integrity failures in import and export flows, and information leakage under exceptional conditions. Unlike functional tests, security tests are adversarial. They use crafted inputs and boundary violating requests and then check whether the observed behavior matches the security expectation.

\paragraph{Baseline setup for Detect, Repair, and Verify.}
Each implementation is initialized in a controlled baseline state in which the full functional test suite passes, establishing a semantically correct reference prior to any security hardening. The security suite is then executed on the same baseline to identify a fixed set of \emph{initially failing} security-test targets ($N$) for that language. These failing targets provide reproducible, triggerable evidence used as the starting point for the Detect--Repair--Verify evaluation. During the experiments, detection produces a structured vulnerability report for each target, repair applies a patch guided by that report, and verification reruns the corresponding security target together with the full functional suite to confirm remediation and detect regressions.

\paragraph{Reducing false passes and false failures in security tests.}
Several design choices are used to improve reproducibility. Whenever possible, each test isolates a single security property. Verdicts rely on two types of evidence: HTTP level signals such as status codes, redirects, headers, and response bodies, and state level signals such as database state, file system state, or object accessibility. For each security family, the suite includes positive controls where authorized actions should succeed and negative controls where unauthorized actions should fail. Test runs are isolated using fresh databases and temporary working directories with fixed seed data. Leakage checks avoid brittle string matching by using stable pattern sets. For payload based tests, reachability is confirmed first so that the relevant code path is exercised before adversarial inputs are applied.

Tables~\ref{tab:test_suites_overview} and~\ref{tab:dataset_test_counts} summarize the structure and scale of the executable benchmark used in this study. Table~\ref{tab:test_suites_overview} characterizes the two complementary test suites in terms of coverage focus, verdict evidence, and their roles in the detect--repair--verify (D--R--V) baseline. The functional suite validates intended workflows under correct authorization and expects successful responses and consistent system states, whereas the security suite issues adversarial probes on high-risk surfaces and expects denial responses and the absence of unsafe side effects. Table~\ref{tab:dataset_test_counts} further quantifies the suite composition across languages, reporting the number of functional tests and initially failing security-test targets per language, as well as overall totals. Table~\ref{tab:security_test_coverage} complements these summaries by breaking down the 26 security targets by \texttt{R01--R10} risk families (OWASP Top~10:2025 mapping), making explicit which vulnerability categories are represented and how the coverage is distributed across them. Together, these tables clarify both the qualitative evaluation signals and the quantitative scale and coverage of the benchmark.

\begin{table}
\centering
\caption{Test-suite structure and verdict signals used in the dataset.}
\label{tab:test_suites_overview}
\begin{tblr}{
  row{1} = {font=\bfseries},
  column{4} = {c},
  cell{1}{1} = {c},
  cell{1}{2} = {c},
  hlines,
  hline{3} = {-}{dotted},
  hline{4} = {-}{0.08em},
}
Suite      & Coverage focus                                                                                                                                                                 & Typical verdict evidence                                                                                                         & Role in D–R–V baseline                           \\
Functional & {Benign workflows under\\~correct authorization~\\(e.g., course/membership,~\\invite join, posting/search,~\\submissions, import/export)}                                      & {HTTP success for~\\allowed actions\\~+~\\expected UI/JSON outputs~\\+~\\expected DB state}                                      & {All tests pass\\~(semantic equivalence)}        \\
Security   & {Adversarial probes of Top~10\\relevant surfaces (e.g., access~\\control/IDOR, injection-style\\~inputs, file/archive handling,\\~integrity of import/export,\\error leakage)} & {HTTP denial (401/403)~\\for boundary violations~\\+~\\absence of unsafe side~\\effects (DB/files)\\~+\\~leakage pattern checks} & {All tests fail~\\(triggerable vulnerabilities)} 
\end{tblr}
\end{table}

\begin{table}
\centering
\caption{Dataset test-suite composition by language. Functional tests validate baseline correctness, while security tests (targets) capture initially failing security requirements used for detect--repair--verify evaluation.}
\label{tab:dataset_test_counts}
\begin{tblr}{
  cells = {c},
  row{1} = {font=\bfseries},
  hlines,
  hline{1,2,6} = {-}{0.08em},
}
Lang   & Functional tests (\#) & Security tests / targets (\#) & Total (\#) \\
Python & 15                    & 7                              & 22         \\
JS     & 8                     & 13                             & 21         \\
PHP    & 14                    & 6                              & 20         \\
Total  & 37                    & 26                             & 63         \\
\end{tblr}
\end{table}

\begin{table}
\centering
\caption{Security-test coverage under the R01--R10 taxonomy (OWASP Top 10:2025 mapping).}
\label{tab:security_test_coverage}
\begin{tabular}{l c}
\toprule
Category & \# Security Tests \\
\midrule
R01 (Access Control / IDOR) & 6 \\
R02 (Security Misconfiguration / Session) & 2 \\
R04 (Cryptographic Failures) & 4 \\
R05 (Injection: SQLi / XSS / CSV) & 6 \\
R08 (Integrity / Upload / Headers) & 4 \\
R09 (Logging \& Monitoring) & 1 \\
R10 (Exception / Error Handling) & 3 \\
\midrule
Total & 26 \\
\bottomrule
\end{tabular}
\end{table}

\subsubsection{Stage~5: Prompt Granularity Design}

This stage defines three prompt granularities used for LLM interaction during the Detect, Repair, and Verify workflow. The underlying applications are unchanged across settings. Only the scope and locality of the context provided to the model vary. This allows studying how granularity affects what issues are detected, how actionable the proposed repairs are, and how the artifacts perform under verification.

\paragraph{Project-level.}
Project-level prompts describe the system as a whole, including the tenant boundary (\texttt{course}), user roles, major modules, and the technology stack. This setting supports broad reasoning across components and is suited for identifying risks that emerge from interactions between features.

\paragraph{Requirement-level.}
Requirement-level prompts present role-scoped requirements and boundary constraints, such as which actions are permitted for a given role and which resources must be protected. This setting emphasizes boundary reasoning and connects potential risks to specific workflows and entry surfaces.

\paragraph{Function-level.}
Function-level prompts focus on a single use case or endpoint. They specify inputs and outputs, preconditions such as authorization, and the relevant resources and data objects. The intent is to encourage localized changes, improve repair actionability, and reduce semantic drift. Effects are evaluated through post-repair security re-testing and function-test regression checks.

The full prompt specifications for vulnerability detection and vulnerability repair, across all three prompt granularities, are publicly available in the accompanying repository\footnote{\url{https://github.com/Hahappyppy2024/EmpricalVDR}}.

\subsection{D--R--V Workflows and Verification-Based Measures}
This section instantiates the workflow in Fig.~\ref{fig:methodology} for evaluation on the multi-language EduCollab benchmark. It first defines the compared workflow conditions (W0--W2) under a matched budget cap, including the bounded iterative setting with iteration limit $K=2$. It then specifies the three prompt granularities used to elicit detection and repair behaviors. Next, it describes the three stages of the pipeline---detection, repair, and verification---with verification grounded in executable functional and security test suites. Finally, it explains how per-target test outcomes and recorded signals are aggregated into the measures used to answer RQ1--RQ3. Results are reported in Section~\ref{Results}.

\subsubsection{Workflow Conditions and Budget Cap}
Three workflow conditions are compared under a matched interaction budget. W0 (\emph{Baseline}) denotes the unrepaired project state. In this state, the functional test suite is required to pass in full, establishing a correct reference implementation for subsequent repairs, while the security test suite is designed to fail, yielding a fixed set of failing security-test targets used for evaluation.

W1 (\emph{Single-pass D--R--V}) performs one detect--repair--verify round per target. For each failing target, a detection prompt is issued once to obtain a repair plan and any referenced files/APIs, followed by a single repair prompt to produce a minimal patch. The patch is then validated by rerunning the target security test and the full functional suite.

W2 (\emph{Bounded iterative D--R--V}) extends W1 with test-grounded feedback under a fixed iteration cap of $K=2$. Each iteration consists of (i) detection and repair conditioned on the current failure evidence and (ii) verification using the same target security test together with the full functional suite. The process stops early when the target security test passes and no functional regression is introduced. If the first repair attempt does not succeed, a second iteration is allowed. In that iteration, the repair prompt is augmented with additional code context requested or implied by the model output (e.g., referenced files, routes, or APIs), provided as file-level snippets to support a revised repair attempt. The workflow terminates when a verified fix is obtained or when the iteration budget is exhausted, keeping W1 and W2 comparable under the same cap.

\subsubsection{Prompt Granularities}
Detection and repair are elicited under three prompt granularities that differ in how much system context is provided to the model. The goal is to separate broad, architecture-level reasoning from role-scoped analysis and from target-driven, test-grounded repair.

\textbf{Project-level prompts} provide a high-level description of EduCollab, including its multi-tenant setting, roles, and major modules and entry points. Under this scope, the model is asked to enumerate plausible vulnerability risks and to summarize them using the \texttt{R01--R10} taxonomy with accompanying localization tags (Layer and Surface). These prompts are used in RQ2 to characterize the breadth and structure of detection reports when the model reasons from a global system description.

\textbf{Requirement-level prompts} narrow the analysis to a specific security requirement under a given role perspective (Admin, Teacher, Assistant, Student, or Unauthenticated). Each requirement-level prompt asks for vulnerability possibilities relevant to that role and requirement, again constrained to \texttt{R01--R10} and the same localization tags. For each model--language setting, five role-scoped responses are collected and summarized in RQ2 to capture both the typical report content (mean across roles) and the variation induced by role conditioning.

\textbf{Function-level prompts} are target-driven and test-grounded. For a failing security-test target, the prompt provides the failing test and the minimal code context needed for repair, and the model is asked to produce a minimal patch for the referenced files only. Function-level prompts are used in W1/W2 to execute the D--R--V pipeline and support RQ1 and RQ3 by producing verifiable outcomes under executable testing.

\subsubsection{Large Language Model Selection}
ChatGPT-5 and GLM-5 are selected as two strong instruction-following models for code-related tasks, enabling cross-model comparison under the same benchmark and test-grounded D--R--V verification protocol (Fig.~\ref{fig:methodolgy}). ChatGPT-5 serves as a widely adopted general-purpose baseline for developer assistance and code generation \cite{openai_gpt4_techreport}. GLM-5 is included as a representative model from the GLM line, providing a contrasting provider/model family to examine whether the observed detection and repair behaviors persist beyond a single ecosystem \cite{zeng2026glm5,zeng2023glm}. Evaluating both models under identical workflow conditions helps disentangle the effects of the D--R--V design and verification feedback from model-specific idiosyncrasies.

\subsubsection{Detection Stage}
The detection stage produces a structured detection report for each model--language setting under a given prompt granularity. To ensure comparability across models and languages, the report is constrained to the \texttt{R01--R10} taxonomy (mapped to OWASP Top~10:2025) and a shared set of localization tags. For each reported risk, the output records (i) the \texttt{R01--R10} family, (ii) the suspected \emph{Layer} (a code-layer cue), and (iii) the suspected \emph{Surface} (an entry-surface cue). The report also includes brief supporting rationale and a minimal, actionable repair plan.

Under project- and requirement-level prompts, the model is asked to enumerate vulnerability possibilities within the given scope using the same \texttt{R01--R10}/Layer/Surface format. Requirement-level detection is issued under five role perspectives (Admin, Teacher, Assistant, Student, and Unauthenticated) to capture how role conditioning changes risk coverage and localization. Results are summarized both by reporting role-wise outputs and by reporting the mean across the five roles for each setting.

Under function-level prompts (used in w0/W1/W2), detection is driven by a failing security-test target. The prompt provides the failing test and the available code snippets, and the model is asked to explain the failure, identify the most relevant file(s) and API/route(s), and propose a concrete repair plan for the subsequent repair stage. These function-level reports serve as the immediate repair guidance and are later evaluated for both repair-actionability and test-grounded reliability.

\subsubsection{Repair Stage}
The repair stage turns the preceding detection output into a concrete patch for each failing security-test target. The repair prompt includes the failing test and selected elements from the detection report, including the reported \texttt{R01--R10} category, the Layer/Surface tags, and any referenced files, routes, or APIs. Keeping this information explicit preserves traceability between what was reported and what was changed, which later supports checks on repair sufficiency and location correctness.

Repairs are produced under a minimal-change constraint. The model is restricted to editing only the file name(s) and code snippets provided in the prompt, is not allowed to introduce new dependencies, and is asked to make the smallest change that satisfies the target security property. If resolving a target requires edits in more than one file, changes are confined to the smallest necessary set of files.

The number of repair attempts follows the workflow condition. W1 produces a single patch based on the initial detection report. W2 allows one additional attempt when the first patch does not pass verification. In the second attempt, additional file-level context is supplied when it was missing in the first round and is requested or implied by the model output (e.g., a specific route handler, middleware, or controller). The second attempt uses the same patch constraints and is conditioned on the verification result from the first attempt.

\section{Results}
\label{Results}

\subsection{RQ1 (Pipeline-level effectiveness).}

\subsubsection{Measures.}
RQ1 compares test-grounded outcomes under matched interaction budgets between single-pass detect--repair--verify (W1) and bounded iterative detect--repair--verify (W2, with cap $K=2$), using the unrepaired project state as baseline (W0). For each model--language setting, evaluation uses a fixed functional test suite and a fixed set of initially failing security-test targets. Table~\ref{tab:rq1_function_level_results} reports the suite sizes (\#Func tests and \#Sec tests/targets) and the numbers of passed tests (Func passed and Sec passed) under each workflow.

Secure-and-correct yield (SCY) is measured at the \emph{target} level as the fraction of initially failing security-test targets that become passing \emph{without} introducing any functional regression under a given workflow. In Table~\ref{tab:rq1_function_level_results}, \#SCY reports the corresponding count of secure-and-correct targets for W0/W1/W2. To quantify the effect of bounded iteration under comparable budgets, Table~\ref{tab:rq1_scy_summary} reports SCY as a rate for W1 and W2 and the improvement $\Delta$SCY = SCY(W2)$-$SCY(W1) under the same cap $K=2$.

\subsubsection{Findings.}

Under matched budgets ($K=2$), bounded iterative detect--repair--verify (W2) consistently improves pipeline-level secure-and-correct yield over both the unrepaired baseline (W0) and the single-pass detect--repair baseline (W1). Table~\ref{tab:rq1_function_level_results} shows that, by construction, all security-test targets fail at W0 across models and languages (\emph{Sec passed}=0 and \#SCY=0), establishing a common starting point for comparison. Moving from W0 to W1 yields non-trivial gains in several settings, but also exposes cases where a single pass fails to achieve any secure-and-correct outcomes (e.g., ChatGPT-5--Python remains at \#SCY=0 in W1). In contrast, W2 produces additional secure-and-correct targets in every model--language setting, indicating that test-grounded feedback in the iterative loop enables progress beyond what is achievable in a single round.

The aggregate effect is summarized in Table~\ref{tab:rq1_scy_summary}. For ChatGPT-5, SCY increases from 0.00/0.54/0.17 in W1 to 0.57/0.77/0.50 in W2 for Python/JS/PHP, corresponding to improvements of +0.57, +0.23, and +0.33, respectively. For GLM-5, SCY rises from 0.43/0.23/0.33 in W1 to 0.86/0.69/0.83 in W2, with improvements of +0.43, +0.46, and +0.50. These gains are not driven by changes in evaluation scope: within each language, the functional and security suite sizes are fixed (Table~\ref{tab:rq1_function_level_results}), and W2 is capped to the same iteration budget ($K=2$) across models.

Table~\ref{tab:rq1_function_level_results} further clarifies why SCY improves. The dominant contribution comes from increased security resolution under W2 (higher \emph{Sec passed}) while functional regressions remain limited. For example, ChatGPT-5--Python improves from 0 to 4 security targets passed when moving from W1 to W2, yielding \#SCY=4 without observed functional loss (\emph{Func passed}=15 throughout). For GLM-5--Python, \emph{Sec passed} increases from 3 to 6, producing \#SCY=6. In JS, both models show large security gains under W2 (ChatGPT-5: 7$\rightarrow$11; GLM-5: 3$\rightarrow$10), but each also exhibits one functional regression under W2 (Func passed drops from 8 to 7), resulting in \#SCY=10 and 9, respectively. PHP shows a similar pattern: ChatGPT-5 improves security passes from 1 to 4 under W2 but incurs one functional regression (14$\rightarrow$13), yielding \#SCY=3; GLM-5 attains \#SCY=5 with no functional loss. Overall, the iterative workflow yields net SCY improvements because the additional security fixes outweigh the small number of functional regressions.

The magnitude of the W2 benefit varies by language and model, suggesting that iteration is most valuable in settings where W1 struggles to localize or repair the vulnerability within a single attempt. The largest gains appear for ChatGPT-5--Python (+0.57) and GLM-5--PHP (+0.50), whereas the smallest gain appears for ChatGPT-5--JS (+0.23), where W1 already achieves relatively high SCY (0.54). This pattern is consistent with a “diminishing returns” effect: when W1 performance is already strong, iteration yields smaller incremental improvements under the same budget cap.

Regarding prompt granularity, the present tables report the function-level condition. Granularity effects can be assessed by repeating the same matched-budget comparison (W2 vs W1, $K=2$) at other prompt scopes and reporting SCY and $\Delta$SCY in the same format as Table~\ref{tab:rq1_scy_summary}. This design isolates the contribution of prompt scope from the iterative budget, enabling a direct test of whether broader (or narrower) prompts amplify or attenuate the benefit of test-grounded iteration.

\begin{table}
\centering
\caption{RQ1 summary: secure-and-correct yield (SCY) under matched budgets, comparing single-pass detect--repair (W1) and bounded iterative detect--repair--verify (W2).}
\label{tab:rq1_scy_summary}
\begin{tblr}{
  cells = {c},
  cell{2}{1} = {r=3}{},
  cell{5}{1} = {r=3}{},
  hlines,
  hline{1,2,5,8} = {-}{0.08em},
  hline{3-4,6-7} = {2-7}{dotted},
}
LLM     & Lang   & Granularity & Budget cap & SCY (W1) & SCY (W2) & $\Delta$SCY (W2–W1) \\
ChatGPT-5 & Python & Function    & $K=\,$2    & 0.00     & 0.57     & +0.57               \\
        & JS     & Function    & $K=\,$2    & 0.54     & 0.77     & +0.23               \\
        & PHP    & Function    & $K=\,$2    & 0.17     & 0.50     & +0.33               \\
GLM-5   & Python & Function    & $K=\,$2    & 0.43     & 0.86     & +0.43               \\
        & JS     & Function    & $K=\,$2    & 0.23     & 0.69     & +0.46               \\
        & PHP    & Function    & $K=\,$2    & 0.33     & 0.83     & +0.50               \\
\end{tblr}
\end{table}

\begin{table}
\centering
\caption{RQ1 function-level results: test-grounded outcomes across workflows. W0 denotes the initial (unrepaired) project state. W1 performs a single detect--repair--verify round, and W2 performs bounded iterative detect--repair--verify. Rows are grouped by language for within-language model comparison.}
\label{tab:rq1_function_level_results}
\begin{tblr}{
  cells = {c},
  cell{2}{1} = {r=6}{},
  cell{2}{2} = {r=3}{},
  cell{5}{2} = {r=3}{},
  cell{8}{1} = {r=6}{},
  cell{8}{2} = {r=3}{},
  cell{11}{2} = {r=3}{},
  cell{14}{1} = {r=6}{},
  cell{14}{2} = {r=3}{},
  cell{17}{2} = {r=3}{},
  hlines = {dotted},
  hline{1-2,8,14,20} = {-}{solid,0.08em},
}
Lang        & LLM       & Workflow         & {\#Func\\tests} & {\#Sec\\tests} & {Func\\passed} & {Sec\\passed} & \#SCY \\
Python      & ChatGPT-5   & W0 Baseline      & 15              & 7              & 15             & 0             & 0    \\
            &           & W1 Single-pass   & 15              & 7              & 15             & 0             & 0    \\
            &           & W2 Iterative     & 15              & 7              & 15             & 4             & 4    \\
            & GLM-5     & W0 Baseline      & 15              & 7              & 15             & 0             & 0    \\
            &           & W1 Single-pass   & 15              & 7              & 15             & 3             & 3    \\
            &           & W2 Iterative     & 15              & 7              & 15             & 6             & 6    \\
JS          & ChatGPT-5   & W0 Baseline      & 8               & 13             & 8              & 0             & 0    \\
            &           & W1 Single-pass   & 8               & 13             & 8              & 7             & 7    \\
            &           & W2 Iterative     & 8               & 13             & 7              & 11            & 10   \\
            & GLM-5     & W0 Baseline      & 8               & 13             & 8              & 0             & 0    \\
            &           & W1 Single-pass   & 8               & 13             & 8              & 3             & 3    \\
            &           & W2 Iterative     & 8               & 13             & 7              & 10            & 9    \\
PHP         & ChatGPT-5   & W0 Baseline      & 14              & 6              & 14             & 0             & 0    \\
            &           & W1 Single-pass   & 14              & 6              & 14             & 1             & 1    \\
            &           & W2 Iterative     & 14              & 6              & 13             & 4             & 3    \\
            & GLM-5     & W0 Baseline      & 14              & 6              & 14             & 0             & 0    \\
            &           & W1 Single-pass   & 14              & 6              & 14             & 2             & 2    \\
            &           & W2 Iterative     & 14              & 6              & 14             & 5             & 5    \\
\end{tblr}
\end{table}

\noindent\textbf{RQ1 takeaway.} Under the same budget cap ($K=2$), bounded iterative D--R--V (W2) consistently improves secure-and-correct yield over single-pass D--R (W1), with gains ranging from +0.23 to +0.57 across model--language settings (Table~\ref{tab:rq1_scy_summary}).

\begin{tcolorbox}[
  colback=gray!10,
  colframe=gray!60,
  boxrule=0.6pt,
  arc=2pt,
  left=6pt,right=6pt,top=4pt,bottom=4pt
]
\noindent RQ1:  Under the same budget cap ($K=2$), bounded iterative D--R--V (W2) consistently improves secure-and-correct yield over single-pass D--R (W1), with gains ranging from +0.23 to +0.57 across model--language settings).
\end{tcolorbox}

\subsection{RQ2 (Detection reliability as repair guidance).}

\subsubsection{Measures.}
RQ2 assesses detection reports from two perspectives: \emph{repair-actionability}, which captures whether a report provides clear and structured guidance for repair, and \emph{test-grounded reliability}, which captures whether reported risks are supported by observable evidence in the benchmark.

\paragraph{Repair-actionable (report usefulness).}
Detection reports are constrained to the \texttt{R01--R10} taxonomy (mapped to OWASP Top~10:2025) and a fixed set of localization tags. Usefulness is summarized with three indicators reported in Tables~13--16: \emph{Vul.\ Types} counts the number of distinct \texttt{R01--R10} families mentioned; \emph{Vul.\ Location Layers} counts distinct code-layer cues; and \emph{Vul.\ Location Surfaces} counts distinct entry-surface cues. Project-level results are reported per model--language setting (Table~13). Requirement-level results are reported as the mean across five role-scoped prompts (Admin, Teacher, Assistant, Student, and Unauthenticated) (Table~14), and role-wise breakdowns are used to show within-setting variation (Tables~15--16).

At the function level, report structure is further summarized using \emph{PlanRate}, \emph{Avg.\#Files}, \emph{Avg.\#APIs}, and \emph{MissingRate} (Table~17). These indicators capture whether a report provides an actionable plan and whether it names the files and APIs needed to carry out a repair.

\paragraph{Reliable (test-grounded correctness).}
Reliability is evaluated against executable security tests under the same model--language setting. A reported \texttt{R01--R10} family is treated as test-grounded when at least one failing security test consistent with that family is observed in the corresponding setting. Downstream verification provides an additional signal for practical reliability: reports are treated as more reliable when they can support a patch that makes the failing security test pass without causing functional regressions under the same workflow constraints (Table~17 and RQ3 outcomes).

\subsubsection{Findings.}
RQ2 is examined from two angles: whether detection reports are \emph{repair-actionable} (i.e., provide clear, localized guidance that can be followed), and whether they are \emph{reliable} under test-grounded evidence. The tables in this section provide complementary views across prompt scope (project vs.\ requirement), role conditioning (five role prompts), and function-level behavior in the detect--repair pipeline.

\paragraph{Repair-actionable (report usefulness).}
Tables~\ref{tab:rq2_project_level_summary}--\ref{tab:rq2_requirement_level_summary} focus on the \emph{content} that a report exposes for repair. At the project level (Table~\ref{tab:rq2_project_level_summary}), both models report broad risk coverage under \texttt{R01--R10}: ChatGPT-5 identifies 9--10 families across languages, while GLM-5 identifies 8--9. Localization cues are also present at this scope (5--7 layers and 5--6 surfaces), but the specificity varies by model--language setting. For example, in JavaScript the project-level report from GLM-5 covers fewer risk families and fewer layer cues (8 families, 5 layers) than ChatGPT-5 (10 families, 6 layers), whereas the gap is smaller in Python and PHP. This indicates that project-level prompts generally elicit wide enumeration, while the strength of localization hints is more setting-dependent.

Requirement-level prompts (Table~\ref{tab:rq2_requirement_level_summary}) make the difference in usefulness clearer. Averaged over the five role-scoped prompts, ChatGPT-5 remains consistently informative across all three languages (8.8--9.2 families; 6.0--6.8 layers; 5.0--5.4 surfaces). In contrast, GLM-5 reports fewer families and fewer location cues (5.2--5.8 families; 4.6--5.2 layers; 3.2--3.4 surfaces). Overall, role-scoped prompting surfaces a larger model gap than project-level prompting, especially for localization cues that are directly needed to guide follow-up fixes.

A zoom-in view illustrates how usefulness varies across roles. In Python (Table~\ref{tab:rq2_requirement_level_by_role_python}), ChatGPT-5 is relatively stable across roles (typically 8--9 families, 6 layers, and 4--6 surfaces), while GLM-5 varies more and provides the weakest surface cues for the unauthenticated role (2 surfaces). This role sensitivity matters for repair guidance because boundary-facing roles are often where entry surfaces and enforcement points must be identified precisely.

\paragraph{Reliable (test-grounded correctness).}
Test-grounded reliability is evaluated by linking report guidance to observable failures in the executable benchmark. In this setting, a report is treated as more reliable when its guidance aligns with failing security-test targets under the same model--language setting and supports verified fixes. Table~\ref{tab:rq3_detection_repair_failure_modes} provides an outcome-based signal through Repair-Success-Rate. Several settings illustrate a gap between structure and reliability: despite PlanRate=1.00, single-pass success can be low (e.g., ChatGPT-5--Python is 0.00 in W1), indicating that the report is followable but may be incorrect in localization or root-cause attribution.

Bounded iteration (W2) reduces this gap in every setting. Repair-Success-Rate increases from W1 to W2 for both models and all languages (e.g., ChatGPT-5--Python: 0.00$\rightarrow$0.57; GLM-5--Python: 0.43$\rightarrow$0.86; GLM-5--JavaScript: 0.23$\rightarrow$0.77; ChatGPT-5--PHP: 0.17$\rightarrow$0.67). Overall, Tables~\ref{tab:rq2_project_level_summary}--\ref{tab:rq2_requirement_level_by_role_python} show what reports contain and how that content varies with prompt scope and role, while Table~\ref{tab:rq3_detection_repair_failure_modes} shows whether those reports function as reliable repair guidance under test-grounded verification.

\begin{table}
\centering
\caption{Project-level detection summary under the \texttt{R01--R10} taxonomy (OWASP Top 10:2025), reported per model--language setting.}
\label{tab:rq2_project_level_summary}
\begin{tblr}{
  cells = {c},
  hlines = {dotted},
  hline{1-2,8} = {-}{solid},
}
LLM     & Lang   & Granularity & {Vul. Types~\\(unique R01–R10)} & {Vul.~\\Location Layers} & {Vul.\\Location Surfaces} \\
ChatGPT-5 & Python & Project     & 10                                &7                          &  6                         \\
GLM-5   & Python & Project     &    9                             & 6                         &  6                         \\
ChatGPT-5 & JS     & Project     &10                                 &6                          & 6                          \\
GLM-5   & JS     & Project     & 8                                & 5                         &6                           \\
ChatGPT-5 & PHP    & Project     &9                                 &  7                        &6                           \\
GLM-5   & PHP    & Project     & 9                                & 6                         &   5                        
\end{tblr}
\end{table}

\begin{table}[t]
\centering
\caption{Requirement-level detection summary under the \texttt{R01--R10} taxonomy (OWASP Top 10:2025). Reported values are the mean across five role-specific prompts (Admin, Teacher, Assistant, Student, and Unauthenticated).}
\label{tab:rq2_requirement_level_summary}
\begin{tblr}{
  cells = {c},
  hlines = {dotted},
  hline{1-2,8} = {-}{solid},
}
LLM     & Lang   & Granularity & {Vul. Types~\\(unique R01--R10)} & {Vul.~\\Location Layers} & {Vul.\\Location Surfaces} \\
ChatGPT-5 & Python & Requirement & 8.8 & 6.0 & 5.4 \\
GLM-5   & Python & Requirement & 5.2 & 4.6 & 3.2 \\
ChatGPT-5 & JS     & Requirement & 9.0 & 6.4 & 5.0 \\
GLM-5   & JS     & Requirement & 5.8 & 4.8 & 3.4 \\
ChatGPT-5 & PHP    & Requirement & 9.2 & 6.8 & 5.2 \\
GLM-5   & PHP    & Requirement & 5.8 & 5.2 & 3.4
\end{tblr}
\end{table}

\begin{table}
\centering
\caption{Requirement-level detection breakdown by role for Python. Each row corresponds to one role-specific prompt response.}
\label{tab:rq2_requirement_level_by_role_python}
\begin{tblr}{
  cells = {c},
  cell{2}{1} = {r=5}{},
  cell{2}{2} = {r=5}{},
  cell{2}{3} = {r=5}{},
  cell{7}{1} = {r=5}{},
  cell{7}{2} = {r=5}{},
  cell{7}{3} = {r=5}{},
  hlines = {dotted},
  hline{1} = {-}{solid},
  hline{2,7,12} = {-}{solid,0.08em},
}
LLM         & Lang       & Granularity     & Role      & {Vul. Types~\\(unique R01–R10)} & {Vul.~\\Location Layers} & {Vul.\\Location Surfaces} \\
ChatGPT-5     & Python     & Requirement     & Admin     & 9                               & 6                        & 5                         \\
            &            &                 & Teacher   & 9                               & 6                        & 6                         \\
            &            &                 & Assistant & 9                               & 6                        & 6                         \\
            &            &                 & Student   & 9                               & 6                        & 6                         \\
            &            &                 & Unauth    & 8                               & 6                        & 4                         \\
GLM-5       & Python     & Requirement     & Admin     & 5                               & 6                        & 3                         \\
            &            &                 & Teacher   & 5                               & 4                        & 4                         \\
            &            &                 & Assistant & 6                               & 4                        & 4                         \\
            &            &                 & Student   & 5                               & 5                        & 3                         \\
            &            &                 & Unauth    & 5                               & 4                        & 2                         
\end{tblr}
\end{table}

\begin{table}
\centering
\caption{Detection-to-repair outcomes and failure-mode distribution. For each model--language--workflow setting, PlanRate / Avg.\#Files / Avg.\#APIs / MissingRate summarize the structure of the detection report over failing security-test targets ($N$). PatchRate and Repair-Success-Rate are measured after the repair--verify step. Failure-mode columns are computed \emph{within unsuccessful targets} ($N_{\text{unsucc}}$). BudgetExceeded applies only to the bounded iterative workflow (W2); W0 entries are not applicable.}
\label{tab:rq3_detection_repair_failure_modes}
\begin{tblr}{
  cells = {c},
  cell{2}{1} = {r=6}{},
  cell{2}{2} = {r=3}{},
  cell{3}{8} = {font=\itshape},
  cell{3}{10} = {font=\itshape},
  cell{4}{8} = {font=\itshape},
  cell{4}{10} = {font=\itshape},
  cell{5}{2} = {r=3}{},
  cell{6}{8} = {font=\itshape},
  cell{6}{10} = {font=\itshape},
  cell{7}{8} = {font=\itshape},
  cell{7}{10} = {font=\itshape},
  cell{8}{1} = {r=6}{},
  cell{8}{2} = {r=3}{},
  cell{9}{10} = {font=\itshape},
  cell{10}{10} = {font=\itshape},
  cell{11}{2} = {r=3}{},
  cell{12}{10} = {font=\itshape},
  cell{13}{10} = {font=\itshape},
  cell{14}{1} = {r=6}{},
  cell{14}{2} = {r=3}{},
  cell{15}{10} = {font=\itshape},
  cell{16}{10} = {font=\itshape},
  cell{17}{2} = {r=3}{},
  cell{18}{10} = {font=\itshape},
  cell{19}{10} = {font=\itshape},
  hlines = {dotted},
  hline{1-2} = {-}{solid},
  hline{8,14,20} = {solid},
}
Lang    & LLM      & Workflow           & {\#\\Security\\Test} & {Plan\\Rate} & {Avg.\#\\Files} & {Avg.\#\\APIs} & {Missing\\Rate}        & {Patch\\Rate} & {Repair-\\Success\\Rate}        \\
Python~ & ChatGPT-5~ & W0 Baseline        & 7                    & –            & –               & –              & –                      & –             & –                               \\
        &          & W1 Single-pass DRV & 7                    & 1.00         & 0.29            & 1.14           & \textit{\textit{0.00}} & 1.00          & \textit{\textit{\textit{0.00}}} \\
        &          & W2 Iterative DRV   & 7                    & 1.00         & 0.29            & 1.14           & \textit{\textit{0.00}} & 1.00          & \textit{\textit{\textit{0.57}}} \\
        & GLM-5    & W0 Baseline        & 7                    & –            & –               & –              & –                      & –             & –                               \\
        &          & W1 Single-pass DRV & 7                    & 1.00         & 1.43            & 1.14           & \textit{\textit{0.00}} & 1.00          & \textit{0.43}                   \\
        &          & W2 Iterative DRV   & 7                    & 1.00         & 1.43            & 1.14           & \textit{\textit{0.00}} & 1.00          & \textit{0.86}                   \\
JS~     & ChatGPT-5  & W0 Baseline        & 13                   & –            & –               & –              & –                      & –             & –                               \\
        &          & W1 Single-pass DRV & 13                   & 1.00         & 1.38            & 1.62           & 0.00                   & 1.00          & \textit{\textit{\textit{0.54}}} \\
        &          & W2 Iterative DRV   & 13                   & 1.00         & 1.38            & 1.62           & 0.00                   & 1.00          & \textit{\textit{\textit{0.85}}} \\
        & GLM-5    & W0 Baseline        & 13                   & –            & –               & –              & –                      & –             & –                               \\
        &          & W1 Single-pass DRV & 13                   & 1.00         & 2.00            & 0.92           & 0.00                   & 1.00          & \textit{\textit{\textit{0.23}}} \\
        &          & W2 Iterative DRV   & 13                   & 1.00         & 2.00            & 0.92           & 0.00                   & 1.00          & \textit{\textit{\textit{0.77}}} \\
PHP~    & ChatGPT-5  & W0 Baseline        & 6                    & –            & –               & –              & –                      & –             & –                               \\
        &          & W1 Single-pass DRV & 6                    & 1.00         & 0.00            & 1.50           & 0.00                   & 1.00          & \textit{\textit{\textit{0.17}}} \\
        &          & W2 Iterative DRV   & 6                    & 1.00         & 0.00            & 1.50           & 0.00                   & 1.00          & \textit{\textit{\textit{0.67}}} \\
        & GLM-5    & W0 Baseline        & 6                    & –            & –               & –              & –                      & –             & –                               \\
        &          & W1 Single-pass DRV & 6                    & 1.00         & 0.00            & 0.67           & 0.00                   & 1.00          & \textit{\textit{\textit{0.33}}} \\
        &          & W2 Iterative DRV   & 6                    & 1.00         & 0.00            & 0.67           & 0.00                   & 1.00          & \textit{\textit{\textit{0.83}}} 
\end{tblr}
\end{table}

\begin{tcolorbox}[
  colback=gray!10,
  colframe=gray!60,
  boxrule=0.6pt,
  arc=2pt,
  left=6pt,right=6pt,top=4pt,bottom=4pt
]
\noindent \textbf{RQ2:} Detection reports are generally repair-actionable (high PlanRate with near-zero MissingRate) but vary in reliability, as reflected by the gap between structured guidance and test-grounded repair success.
\end{tcolorbox}

\subsection{RQ3 (Repair trustworthiness and dominant failure modes).}

\subsubsection{Measures.}
RQ3 evaluates repair trustworthiness under test-grounded verification. The unit of analysis is a failing security-test target. For each model--language--workflow setting, $N$ denotes the number of initially failing security tests (targets). W0 is the unrepaired baseline, W1 is single-pass detect--repair--verify, and W2 is bounded iterative detect--repair--verify under the same budget cap ($K=2$).

Repair effectiveness is summarized by \emph{Fixed}, the number of targets that become passing after the repair--verify step, and \emph{FixRate} $=$ Fixed$/N$. Repair trustworthiness further considers side effects on functional correctness. \emph{FuncReg} counts functional regressions introduced by the patch, measured as the number of previously passing functional tests that fail after patch application and verification.

Dominant failure modes are analyzed within unsuccessful targets, where $N_{\text{unsucc}} = N - \textit{Fixed}$. Under this setup, a target is \emph{NotFixed} when it remains failing after the repair--verify step. Among these unsuccessful targets, \emph{WrongLocalization} captures cases where the recorded localization verdict indicates that the repair attempt did not target the correct location (\texttt{location\_right}=false). WrongLocalization is summarized by \emph{WrongLoc} (count) and \emph{WrongLocRate} $=$ WrongLoc$/N_{\text{unsucc}}$. More fine-grained failure categories such as semantic drift or newly introduced security issues require additional evidence beyond the current recorded signals and are not quantified in this study.

\subsubsection{Findings.}
Table~\ref{tab:rq3_detection_repair_failure_modes} summarizes repair outcomes under test-grounded verification. The unrepaired baseline (W0) resolves none of the failing security-test targets in any setting (FixRate$=0.00$).

Across languages and models, bounded iterative D--R--V (W2, $K=2$) consistently yields higher FixRate than the single-pass workflow (W1). For ChatGPT-5, FixRate increases from 0.00 to 0.57 in Python and from 0.54 to 0.85 in JavaScript; for GLM-5, FixRate increases from 0.43 to 0.86 in Python and from 0.23 to 0.77 in JavaScript. Similar gains are observed in PHP. These results indicate that verification feedback in bounded iteration converts additional failing targets into passing ones under the same budget cap.

Functional side effects are uncommon but not absent. No functional regressions are observed in Python for either model. In contrast, one regression is observed under W2 in JavaScript for both models, and one regression is observed under W2 in PHP for ChatGPT-5. Overall, iterative repair improves verified security fixes while keeping functional regressions largely contained in this benchmark.

Dominant failure modes are assessed within unsuccessful targets ($N_{\text{unsucc}}$), where the security test remains failing after verification (\emph{NotFixed}). Table~\ref{tab:rq3_failure_mode_wrongloc} shows that most remaining failures are associated with incorrect repair targeting (\emph{WrongLocalization}, \texttt{location\_right}=false). WrongLocRate is high across settings, often exceeding 0.75 within $N_{\text{unsucc}}$, indicating that unsuccessful repairs are more frequently limited by inaccurate localization than by the absence of a patch. More fine-grained failure characterization beyond localization (e.g., semantic drift or newly introduced security issues) would require additional evidence and is not quantified here.

\begin{table}
\centering
\caption{Repair outcomes and side effects under test-grounded verification. For each model--language--workflow setting, $N$ denotes the number of initially failing security tests (targets). Fixed and FixRate summarize the number and fraction of targets repaired after the repair--verify step, while FuncReg counts functional regressions introduced by the patch.}
\label{tab:rq3_detection_repair_failure_modes}
\begin{tblr}{
  cells = {c},
  cell{2}{1} = {r=6}{},
  cell{2}{2} = {r=3}{},
  cell{5}{2} = {r=3}{},
  cell{8}{1} = {r=6}{},
  cell{8}{2} = {r=3}{},
  cell{11}{2} = {r=3}{},
  cell{14}{1} = {r=6}{},
  cell{14}{2} = {r=3}{},
  cell{17}{2} = {r=3}{},
  hlines = {dotted},
  hline{1-2} = {-}{solid},
  hline{8,14,20} = {solid},
}
Lang    & LLM      & Workflow           & {\# Security\\Test} & Fixed & FixRate & FuncReg \\
Python~ & ChatGPT-5~ & W0 Baseline        & 7                   & 0     & 0.00    & –       \\
        &          & W1 Single-pass DRV & 7                   & 0     & 0.00    & 0       \\
        &          & W2 Iterative DRV   & 7                   & 4     & 0.57    & 0       \\
        & GLM-5    & W0 Baseline        & 7                   & 0     & 0.00    & –       \\
        &          & W1 Single-pass DRV & 7                   & 3     & 0.43    & 0       \\
        &          & W2 Iterative DRV   & 7                   & 6     & 0.86    & 0       \\
JS      & ChatGPT-5  & W0 Baseline        & 13                  & 0     & 0.00    & –       \\
        &          & W1 Single-pass DRV & 13                  & 7     & 0.54    & 0       \\
        &          & W2 Iterative DRV   & 13                  & 11    & 0.85    & 1       \\
        & GLM-5    & W0 Baseline        & 13                  & 0     & 0.00    & –       \\
        &          & W1 Single-pass DRV & 13                  & 3     & 0.23    & 0       \\
        &          & W2 Iterative DRV   & 13                  & 10    & 0.77    & 1       \\
PHP~    & ChatGPT-5  & W0 Baseline        & 6                   & 0     & 0.00    & –       \\
        &          & W1 Single-pass DRV & 6                   & 1     & 0.17    & 0       \\
        &          & W2 Iterative DRV   & 6                   & 4     & 0.67    & 1       \\
        & GLM-5    & W0 Baseline        & 6                   & 0     & 0.00    & –       \\
        &          & W1 Single-pass DRV & 6                   & 2     & 0.33    & 0       \\
        &          & W2 Iterative DRV   & 6                   & 5     & 0.83    & 0       
\end{tblr}
\end{table}

\begin{table}
\centering
\caption{Dominant failure mode under verification. For each model--language--workflow setting, $N_{\text{unsucc}}$ denotes the number of security-test targets that remain failing after the repair--verify step. WrongLoc counts targets with incorrect repair localization (\texttt{location\_right}=false) within $N_{\text{unsucc}}$, and WrongLocRate is computed as WrongLoc$/N_{\text{unsucc}}$.}
\label{tab:rq3_failure_mode_wrongloc}
\begin{tblr}{
  cells = {c},
  cell{2}{1}  = {r=4}{},  
  cell{2}{2}  = {r=2}{},  
  cell{4}{2}  = {r=2}{},  
  cell{6}{1}  = {r=4}{},  
  cell{6}{2}  = {r=2}{},  
  cell{8}{2}  = {r=2}{},  
  cell{10}{1} = {r=4}{},  
  cell{10}{2} = {r=2}{},  
  cell{12}{2} = {r=2}{},  
  hlines = {dotted},
  hline{1-2} = {-}{solid},
  hline{6,10,14} = {-}{solid,0.08em}, 
}
Lang   & LLM     & Workflow           & $N_{\text{unsucc}}$ & WrongLoc & WrongLocRate \\
Python & ChatGPT-5 & W1 Single-pass DRV  & 7  & 6 & 0.86 \\
       &         & W2 Iterative DRV    & 3  & 2 & 0.67 \\
       & GLM-5   & W1 Single-pass DRV  & 4  & 3 & 0.75 \\
       &         & W2 Iterative DRV    & 1  & 0 & 0.00 \\
JS     & ChatGPT-5 & W1 Single-pass DRV  & 6  & 6 & 1.00 \\
       &         & W2 Iterative DRV    & 2  & 2 & 1.00 \\
       & GLM-5   & W1 Single-pass DRV  & 10 & 9 & 0.90 \\
       &         & W2 Iterative DRV    & 3  & 3 & 1.00 \\
PHP    & ChatGPT-5 & W1 Single-pass DRV  & 5  & 5 & 1.00 \\
       &         & W2 Iterative DRV    & 2  & 2 & 1.00 \\
       & GLM-5   & W1 Single-pass DRV  & 4  & 3 & 0.75 \\
       &         & W2 Iterative DRV    & 1  & 0 & 0.00 \\
\end{tblr}
\end{table}

\begin{tcolorbox}[
  colback=gray!10,
  colframe=gray!60,
  boxrule=0.6pt,
  arc=2pt,
  left=6pt,right=6pt,top=4pt,bottom=4pt
]
\noindent \textbf{RQ3:} Iterative D--R--V (W2, $K=2$) substantially increases verified fixes with rare functional regressions, while the dominant remaining failure mode under verification is \emph{WrongLocalization}---most unfixed targets are associated with incorrect repair targeting (WrongLocRate typically 0.67--1.00 within $N_{\text{unsucc}}$).
\end{tcolorbox}

\section{Discussion}
\label{disscusion}
This section connects the results of RQ1--RQ3 and reflects on what they imply for LLM-assisted vulnerability management.

\subsection{Pipeline-level behavior under iteration}
Iteration turns repair from a one-shot patch into a series of small changes, each checked by tests. In practice, the loop is only useful when the verification signals are readable: security tests indicate whether the targeted exploit path is no longer reachable, and functional tests indicate whether expected behavior still holds. A bounded budget makes the loop usable: it avoids open-ended tweaking, while still allowing early stopping once additional edits no longer improve verified outcomes.

Across settings, two ingredients largely determine whether iteration helps: the sharpness of the detection guidance and the size of the edits it triggers. When the report points to a concrete location and the repair stays local, repeated cycles often make measurable progress and keep functionality stable. When the report is vague, the repair tends to spread across files and interfaces; later iterations then spend effort undoing side effects or chasing new symptoms (regressions, contract drift) rather than removing the original cause. This is why iteration cannot be discussed in isolation: verification, prompt granularity, and edit locality move together.

\subsection{Detection as repair guidance}
Detection is the handoff from “something is wrong” to “change this code.” A report is useful as repair guidance only when it can be acted on directly. Naming a category is not enough; the report needs (i) where the issue occurs, (ii) how to trigger it, and (iii) what the secure behavior should look like. When these elements are present—e.g., a specific endpoint, parameters, and an access boundary—the repair step is more likely to touch the right component and produce a change that the tests can confirm.

Unhelpful reports fail in familiar ways. Some stay at the warning level and never commit to a code path. Others mix correct labels with the wrong story, describing a scenario that does not match the implemented workflow. In both cases, repairs become guesswork: changes land in the wrong place, or “hardening” spreads broadly and breaks unrelated behavior. A practical takeaway is to treat detection outputs as hypotheses, not decisions. The most reliable reports look like small, testable claims with a location, a trigger, and an expected outcome.

\subsection{Repair trustworthiness under verification}
Verification is the main basis for trusting a repair. Security checks provide evidence that the reported failure is no longer reproducible under the tested threat model; functional tests provide evidence that the intended workflows still work. Both are necessary because many security fixes touch control flow, validation, or authorization, and those are exactly the areas that can silently change behavior.

After repair, outcomes often cluster into three cases. The best case is a clean win: the security failure disappears and functionality remains intact. A second case improves security signals but breaks functionality, often because the fix over-restricts access, tightens input checks beyond what legitimate requests require, or changes an interface used elsewhere. The third case looks plausible in code but does not move the verified outcomes: the vulnerable path remains reachable, the fix is incomplete, or the failure is displaced rather than resolved. In other words, a patch should be treated as provisional until tests confirm both mitigation and stability.

\subsection{Implications for practice and benchmarking}
For practitioners, the results point to a straightforward workflow rule: let verification drive the loop. Running security and functional tests after each repair attempt turns iteration into evidence-based hardening instead of speculative editing. Prompt granularity also affects day-to-day reliability. Project-level context helps reason about cross-module workflows and access boundaries; function-level context naturally limits the blast radius of edits. Requirement-level prompts often sit in between: they anchor the repair in roles and protected resources, but still leave room for ambiguity in localization.

For benchmarking, the study supports project-level, test-grounded evaluation. Runnable projects with executable functional and security tests make secure-and-correct outcomes observable and make failures easier to attribute to concrete behaviors. Keeping a shared functional specification across implementations also helps maintain comparable semantics in cross-language settings. Extending the benchmark would mainly be a matter of coverage: richer features, more vulnerability classes, and matching tests with stable oracles.

\section{Threats to Validity}
\label{threats}

\subsection{Internal Validity}
Internal validity asks whether the observed differences come from the workflow variants, rather than from the experimental setup.

\textbf{Test-grounded outcomes and oracle adequacy.}
Secure-and-correct yield is assessed with executable oracles: functional tests for correctness, and targeted security tests plus post-repair re-detection for security. The results therefore depend on what the suites cover. Functional tests may miss corner cases or implicit requirements, and security tests only reflect the threat patterns they encode. To keep comparisons fair, the same projects, test suites, and verification procedure are used across all variants and after every repair iteration.

\textbf{Attributing newly introduced security failures.}
It is hard to quantify newly introduced security failures when the baseline security pass rate is near zero, because there is limited signal to detect additional degradation after repair. In such cases, the analysis focuses on fix success (improvements over baseline) and functional regressions captured by functional tests. This avoids over-interpreting an uninformative baseline, but it also limits claims about how often repairs introduce new security issues beyond what re-detection and observed test failures can show.

\textbf{Non-determinism and procedural effects.}
LLM outputs and automated detection can vary with randomness and seemingly minor prompt or execution differences. To reduce run-to-run artifacts, all variants are evaluated under matched budget limits, bounded iterations, and the same stopping rules, while keeping the test suites fixed. Some instance-level variance remains and is treated as part of the workflow behavior being measured.

\subsection{External Validity}
External validity concerns how far the findings may transfer beyond the studied projects, prompts, and test suites.

\textbf{Scope of the benchmark projects.}
The benchmark contains runnable web-application projects with functional and security tests, enabling project-level verification. The findings should not be assumed to carry over to other domains (e.g., systems software, mobile/embedded applications, or smart contracts). Likewise, the security conclusions are tied to the vulnerability patterns exercised by the current security tests and the detectors used for re-detection.

\textbf{Prompt granularity.}
Prompting is varied at the project-, requirement-, and function-level, reflecting common usage modes from whole-project generation to feature implementation and localized patching. Other interaction styles—such as long multi-turn co-design, tool-augmented planning, or large refactoring-oriented prompts—are not covered and may lead to different dynamics.

\textbf{Languages, frameworks, and ecosystems.}
The benchmark spans multiple languages and web stacks, reducing dependence on a single ecosystem. Still, results may change under different frameworks, dependency landscapes, or security-critical runtimes where common vulnerability forms and verification oracles differ.

\textbf{Evolving functionality and test coverage.}
The functional and security tests capture the expected behavior and threat patterns of the current project snapshot. As features grow or new vulnerability classes are added, both suites must evolve to maintain coverage. Extending the benchmark with richer functionality, broader vulnerability classes, and matching tests is likely to improve generality and support more fine-grained analysis.

\section{Conclusion}
\label{conclusion}

This work positions LLM-assisted coding within vulnerability management and operationalizes security hardening as a test-grounded, iterative detect--repair--verify workflow. It introduces a project-level benchmark dataset for end-to-end evaluation, comprising runnable web-application projects paired with executable functional tests and targeted security tests, and supporting three prompt granularities (project-, requirement-, and function-level). Using this benchmark, the study evaluates pipeline-level secure-and-correct yield under comparable budgets, assesses the reliability and repair-actionability of detection reports, and characterizes the trustworthiness of repaired outputs under verification, including dominant failure modes. Future work will extend the benchmark with richer functionality, broaden coverage to additional vulnerability classes, and add corresponding tests to enable more comprehensive and fine-grained evaluation.

\section*{Declarations}

\subsection*{Funding}
This research did not receive any external funding.

\subsection*{Ethical approval}
Not applicable.

\subsection*{Informed consent}
Not applicable.

\subsection*{Author Contributions}
Cheng Cheng: Idea, Methodology, Data collection,and Writing—original draft, review and editing.\\

\subsection*{Data Availability Statement}
The replication package is available at: \texttt{https://github.com/Hahappyppy2024/EmpricalVDR}.

\subsection*{Conflict of Interest}
The authors declare that they have no competing interests.

\subsection*{Clinical trial number}
Clinical trial number: not applicable.

\bibliographystyle{spbasic}      

\bibliography{ref}   

\end{document}